\documentclass[10pt,letterpaper]{article}
\usepackage[top=0.85in,left=1.0in,footskip=0.75in,marginparwidth=2in]{geometry}

\usepackage[utf8]{inputenc}

\usepackage{cite}

\usepackage{nameref,hyperref}

\usepackage[right]{lineno}

\usepackage{microtype}

\usepackage{cleveref}		
\usepackage{graphicx}		
\usepackage{placeins}
\usepackage[utf8]{inputenc}
\usepackage{siunitx}
\usepackage[format=default,indention=.5cm]{caption}

\usepackage{changepage}

\usepackage[aboveskip=1pt,labelfont=bf,labelsep=period,singlelinecheck=off]{caption}

\makeatletter
\renewcommand{\@biblabel}[1]{\quad#1.}
\makeatother

\usepackage{lastpage,fancyhdr,graphicx}
\usepackage{epstopdf}
\fancyheadoffset[L]{2.25in}
\fancyfootoffset[L]{2.25in}

\usepackage{color}

\definecolor{Gray}{gray}{.25}

\usepackage{graphicx}

\begin{document}
\vspace*{0.35in}

\begin{flushleft}
{\Large
\textbf\newline{Structured illumination microscopy using a photonic chip}
}
\newline
\\
\O ystein I. Helle,
Firehun T. Dullo,
Marcel Lahrberg,
Jean-Claude Tinguely

and Balpreet S. Ahluwalia\textsuperscript{*},\\
\bigskip
UiT -The Arctic University of Norway, Department of Physics and Technology, Klokkargårdsbakken 35, 9019 Tromsø, Norway\\

\bigskip
*balpreet.singh.ahluwalia@uit.no

\end{flushleft}

\subsection*{Abstract}

Structured illumination microscopy (SIM) enables live cell, three-dimensional super-resolution imaging of 
sub-cellular structures at high speeds. SIM  uses sophisticated optical systems to generate pre-determined excitation light patterns, and advanced reconstruction algorithms to enhance the optical resolution by up to a factor of two. The optical set-up of SIM relies on free-space optics like gratings or spatial light modulators to generate the light patterns, and a high numerical aperture objective lens to project the pattern on the sample. These arrangements are prone to miss-alignment, often with high initial and maintenance costs, and with the final resolution-enhancement being limited by the numerical aperture of the collection optics. Here, we present a photonic-chip based total internal reflection fluorescence (TIRF) SIM that greatly reduces the complexity of the optical setup needed to acquire TIRF-SIM images. This is achieved by taking out the light generation and delivery from the microscope and transferring it to the photonic-chip. Thus, the conventional glass slide is replaced by a planar photonic chip that both holds and illuminates the specimen. The integrated waveguide chip is used to create a standing wave interference pattern, which illuminates the sample via evanescent fields. The phase of the interference pattern is controlled by the use of thermo-optical modulation, leaving the footprint of the light illumination path for the SIM system to around 4 by 4 cm$^2$. Furthermore, we show that by the use of the photonic-chip technology, the resolution enhancement of SIM can be increased above that of the conventional approach. In addition, by the separation of excitation and collection light paths the technology opens the possibility to use low numerical objective lenses, without sacrificing on the SIM resolution. Chip-based SIM represents a simple, stable and affordable approach, which could enable widespread penetration of the technique and might also open avenues for high throughput optical super-resolution microscopy.


\subsection*{Introduction}
Optical microscopy is limited by diffraction effectively limiting the achievable resolution to around 250 nm laterally, and 500 nm axially \cite{RN660}. The advent of super resolution microscopy, commonly know as nanoscopy, has proven the ability to trick the diffraction limit extending the lateral resolution of the microscope down towards just a few nanometers. One group of methods utilize photo-chemistry manipulation of fluorophores to achieve increased resolution. These techniques include stimulated emission depletion (STED) microscopy \cite{RN661} and single molecule localization microscopy (SMLM) techniques such as direct stochastic optical reconstruction microscopy (\textit{d}STORM)\cite{RN455} and photo-activated localization microscopy (PALM)\cite{RN662}. These techniques offers very good resolution below 10 nm in some cases, but at the sacrifice of temporal resolution, especially over large fields of views. Another branch of methods offer resolution enhancement with less control of the fluorophore chemistry, but relies to a high degree on computational algorithms and fluctuating fluorescence emission to achieve better resolution. These techniques include methods such as super-resolution optical fluctuation imaging (SOFI)\cite{RN526}, entropy based super-resolution imaging (ESI)\cite{RN528} and multiple signal classification algorithm (MUSICAL)\cite{RN530}.

Although the above mentioned techniques give good resolution, the need for manipulating the chemistry of the fluorophore makes them difficult to use, and the low temporal resolution limits its use to fixed samples in most cases. On the other hand, structured illumination microscopy (SIM)\cite{RN663} works for most bright fluorophores, and is inherently fast being a widefield technique. In SIM, a sinusoidal grid pattern illuminates the sample, and the fluorescent emission is captured on a camera. The sinusoidal excitation light combines with the objects via a multiplication, and when convolved through the objective lens the observed fluorescence emission takes the form of a Moirè pattern. This pattern represents a mix of frequencies from both the excitation light and the object, with earlier unresolved information becoming accessible. By using a known spatial frequency in the sinusoidal excitation light, together with phase shifts, the earlier unresolved content can be extracted by the SIM reconstruction algorithm. For improving the resolution along one axis, three images must be acquired using equidistant phase steps of the sinusoidal excitation pattern. For isotropic resolution, the process must be repeated for 3 orientations of the excitation pattern, for a total of 9 images in case of 2D. SIM does not rely on stringent manipulation of the fluorescence, as was the case for the other methods. Since it only needs 9 images to create a super-resolution image the method is fast, making it the most popular method for high temporal resolution nanoscopy imaging. However, the generation of the necessary SIM patterns are usually done using free-space optics. This makes for a complicated optical setup with expensive components to both generate and maintain the necessary pattern orientation, phase steps and polarization state. The resulting SIM microscopes are thus both bulky and expensive, and prone to misalignment needing highly qualified personnel which adds to the maintenance costs. Furthermore, the resolution of conventional linear SIM is limited to 2x over the diffraction limit given by the numerical aperture (N.A.) of the imaging objective lens, a consequence of the SIM pattern being projected via the objective lens. As both illumination and collection light paths are coupled, the use of low N.A. and low magnification objective lenses to image large areas for high-throughput imaging results in much adverse optical resolution.

\paragraph{SIM on a chip.}
In this work we propose a method that will allow any microscope to gather total internal reflection fluorescence (TIRF) SIM images using a mass-producible, and disposable photonic integrated circuit (PIC) chip. In chip-based SIM (cSIM), a low cost upright microscope can gather images of specimen placed directly on top of a planar waveguide  surface. The top layer of the integrated chip consists of a series of optical waveguides that creates the necessary excitation patterns required for SIM. Chip-based microscopy relies on total internal reflection inside the planar waveguides \cite{RN659,RN520,RN694}, where the excitation light is available as an evanescent field exponentially decaying from the surface of the chip. By using an opposing pair of waveguides a standing wave interference pattern can be generated on top of the waveguide surface. This standing wave interference pattern can be harnessed for SIM imaging. This makes for a simplified optical setup as laser light will be coupled on to the waveguides via a fiber without the need for free-space bulky equipment to generate and deliver the light patterns to the sample. By illuminating the sample via waveguides, the excitation and collection light paths are decoupled which alleviates the problem of refractive index oil mismatching as often arises in objective based SIM. Another advantage of cSIM is the decoupling of the illumination and the collection light paths. Thus, irrespective of the imaging objective lens, the SIM illumination pattern and fringe spacing generated by the waveguide is maintained. This opens the opportunity to use low N.A. and low magnification objective lenses to image large areas with uncompromised optical resolution defined by the fringe spacing of the illumination pattern. Moreover, by using an optical waveguide made of high-refractive index material (e.g. n=2.1) the fringe spacing of the standing wave pattern is much smaller than what can be achieved by high N.A. oil immersion objective lens (N.A.=1.49).   
\begin{figure}[ht]
\centering
\includegraphics[height=9cm]{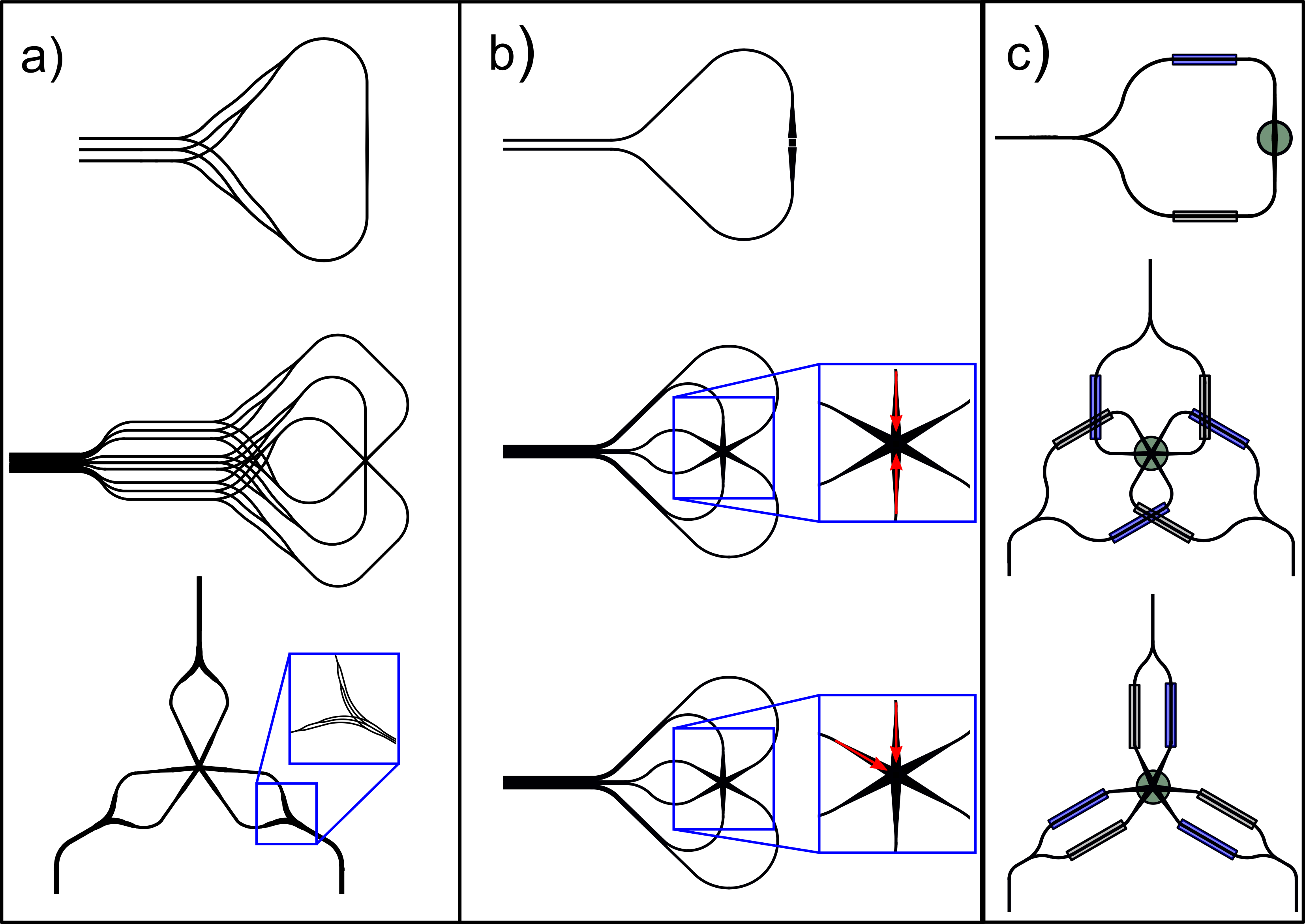}
\caption{Chip-SIM designs shown for 1D SIM with 180$^\circ$ interference (top panel), 2D SIM with 180$^\circ$ interference (middle panel), and 2D SIM with 50/60$^\circ$ interference (bottom panel). a) Phase modulation using fixed optical path-length differences Incorporated by having three different waveguides coming in for each orientation of the SIM pattern. b) Shows cSIM structures for light splitting off-chip, e.g. by using a 50:50 fibre split. Thus, two input waveguides must be used to generate one interference pattern. c) Splitting the light using a y-branch on-chip and changing the phase by using a thermo-optical element located on one of the two waveguide arms. The phase change is the realized by applying a voltage to the resistive element}
\label{Designs}
\end{figure}

SIM imaging relies on accurate phase control of the standing wave interference pattern. In cSIM this can be achieved in several ways. Since the method relies on interference, the laser light must be split in two individually propagating beams. This can be done either on-chip by using a waveguide y-branch; or off-chip by using two waveguides simultaneously, each of them incident with laser light. The accurate phase control of the resulting standing wave can be achieved using different methods. Here, we have investigated 3 different possibilities. For light splitting on-chip in a y-branch, two methods was developed. Figure \ref{Designs}(a), top panel, show a 1D SIM structure where the phase control is realized by using different optical path-lengths for the three phase-steps needed for SIM. In this case, three different structures with y-branches ends up in the same waveguide, each with different optical path lengths but with the same frequency and pattern orientation. Using this design we need three waveguides to have three discrete phase steps for the SIM imaging. For light splitting outside the chip two input waveguides are used for a 1D SIM structure as shown in Fig. \ref{Designs}(b), the phase-stepping can be achieved by manipulating the phase of one of the input light paths, e.g. by using an optical fibre phase modulator or by changing the path-length in one of the arms, e.g by heating one arm in the fiber split. Figure \ref{Designs}(c), top panel, shows on-chip dynamic phase stepping mechanism requiring only one waveguide structure. Here, the phase stepping is achieved by incorporating a thermo-optical heating element on top of one of the waveguide arms. By resistive heating of the buffer material, the local refractive index is changed by thermal expansion, which induces a phase-shift in the SIM pattern at the imaging region (indicated with a circle).

\begin{figure}[ht]
\centering
\includegraphics[width=0.9\linewidth]{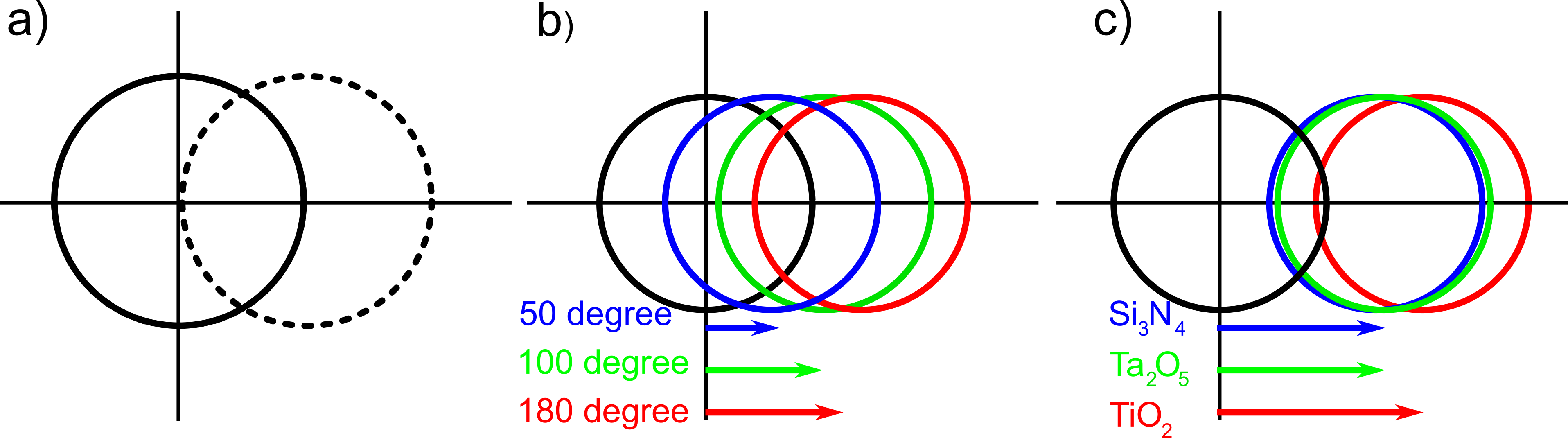}
\caption{a) The solid circle represents the diffraction limited OTF of a microscope. By using conventional SIM the OTF is shifted in frequency space allowing for higher frequencies to be resolved. The dotted circle represent the shifted OTF in frequency space, enabling  frequency content within a maximum of 2*N.A. to be resolved.  b) The frequency shift for chip-based SIM is independent of the imaging objective lens, but governed by the angle of interference and the refractive index of the waveguide material itself. Here, the black circle represents the OTF given by the N.A. of the objective lens, and the colored circles show the cSIM shifts towards higher frequencies for waveguides interfering at angles of 50$^\circ$, 100$^\circ$ and 180$^\circ$, where 180$^\circ$ will provide the best resolution for SIM. c) For higher refractive index materials like Ta$_2$O$_5$ and TiO$_2$, the shift in frequency space can further be increased. Here shown for 180$^\circ$ interference angle.}
\label{OTF}
\end{figure}

The resolution of cSIM is governed by the fringe-spacing of the standing wave interference pattern in addition to the N.A. of the collection optics. By interfering waveguides with changing angles $\theta$ the fringe spacing $f_s$ can be changed as
\begin{equation}
f_s=\frac{\lambda_{ex}}{2n_{eff}\sin \theta/2},
\label{fringespaceing}
\end{equation}
where $n_{eff}$ is the effective refractive index of the guided mode. Generating the SIM pattern with a planar waveguide thus contributes $n_{eff}$ in to the achievable resolution $\Delta_{xy}$ of the method, so the equation describing the theoretical best resolution for cSIM takes the form of
\begin{equation}
\Delta_{xy}=\frac{\lambda_{ex}}{2(N.A. + n_{eff}\sin \theta/2)},
\label{resolution}
\end{equation}
where N.A. is the numerical aperture of the imaging objective lens. Since $n_{eff}$ is larger than N.A. the resolution of the method can exceed that of conventional SIM, which would have $4N.A.$ in the denominator. The resolution enhancement theoretically possible for cSIM is visualized in Fig. \ref{OTF}. Figure \ref{OTF}(a) show the conventional SIM case, where the solid circle indicates the frequency space cut-off (OTF support) of the collection optics. The dotted circle indicates the maximum theoretical shift towards higher frequency content (i.e. increased resolution). The diameter of the solid circle is determined by the numerical aperture of the imaging objective lens and the shift of the dotted circle (resolution enhancement) is given by the smallest fringe period of the illumination pattern. For conventional SIM, the maximum extension of the OTF is 2X. Figure \ref{OTF}(b) show the same situation for cSIM for different interfering angles. By using a high refractive index material, the light can be tightly confined inside the waveguide. The spatial frequency of the interference fringes formed on top of a waveguide can be made higher than for fringes formed in free space, due to the higher refractive index and an angle of 180$^\circ$ between the interfering beams. High effective refractive index of n$_{eff}$=1.7 or n$_{eff}$=1.8 can be obtained for waveguides made of high refractive index material such as silicon nitride (Si$_3$N$_4$) or tantalum pentoxide (Ta$_2$O$_5$), respectively. By using materials of even higher refractive index, e.g. Titanium oxide (TiO$_2$), the cSIM resolution can be further increased. Thus harnessing the high-refractive index of the waveguide material and by interfering angles of 100$^\circ$ and 180$^\circ$ the OTF support is extended above the 2X possible in conventional SIM. Figure \ref{OTF}(c) shows how the choice of different waveguide materials can further enhance the resolution by extending the OTF.

\subsection*{Simulation of waveguide based interference patterns for SIM}
\begin{figure}
		\centering
		\includegraphics[width=0.9\linewidth]{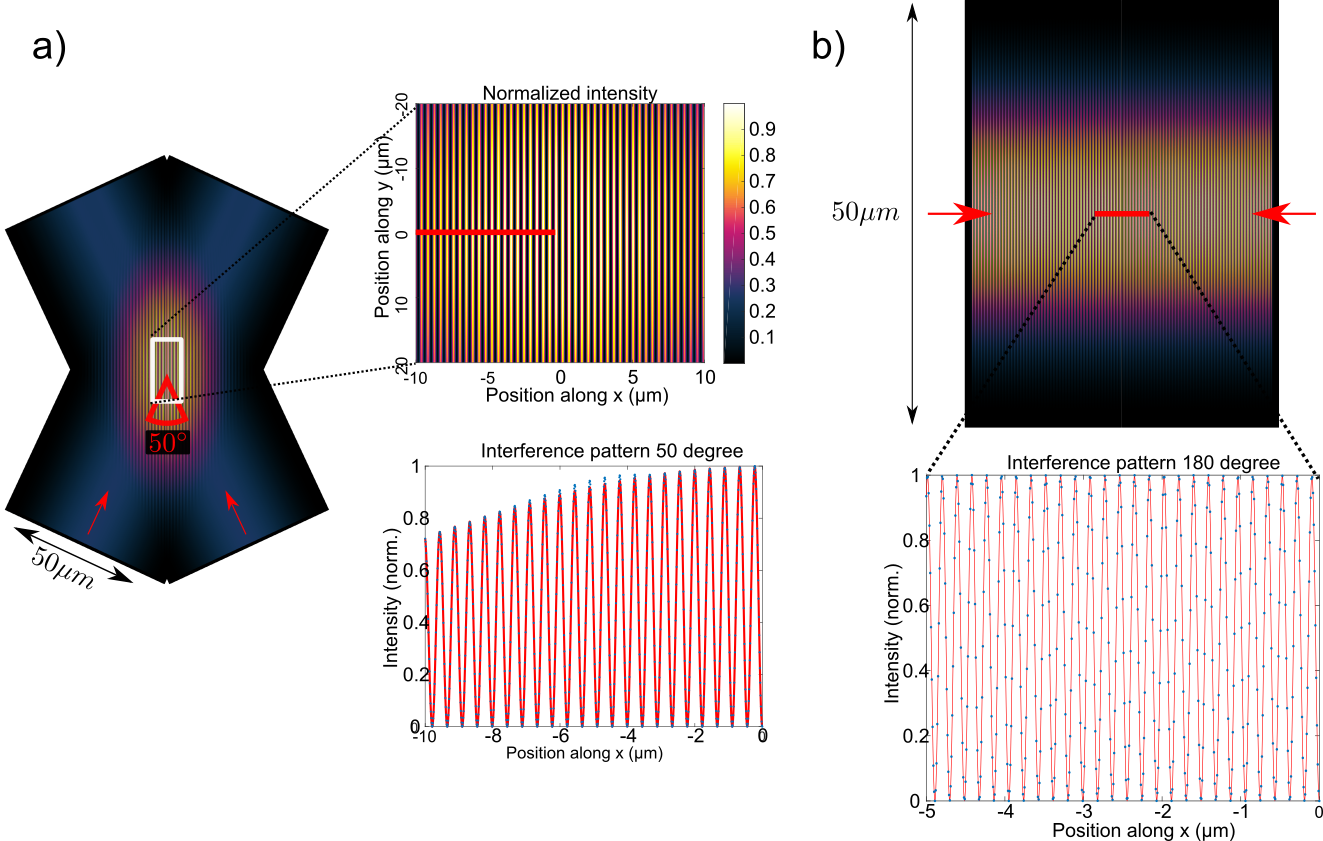}
		\caption[Fringes on waveguide]{Numerical simulations of the proposed waveguide geometries to generate the illumination patterns for cSIM. The configurations of two waveguides crossing at an angle of \SI{50}{\degree} (a) generates a pattern with a periodicity of \SI{445}{\nano\meter}. The couterpropagating waveguides (b) create a pattern with a period of  \SI{188}{\nano\meter}.}
		\label{fig:exportinterferencepatterns}
	\end{figure}

The cSIM interference patterns were simulated using finite element method in Comsol Multiphysics. Figure \ref{fig:exportinterferencepatterns}(a) shows two waveguides crossing at an angle of \SI{50}{\degree}. The red arrows indicate the directions of the straight waveguides along which the light propagates. Since the presented model is limited to a two-dimensional representation of the actual waveguide, the effective refractive indices along the third dimension have been calculated approximating those regions as planar waveguides \cite{Okamoto2006}. Given the fundamental transversal electric (TE) mode being exited in each waveguide, the resulting intensity distribution shows the intended sinusoidal structure as well as a Gaussian envelope due to the shape of the waveguide modes. The mode index in the simulated Si$_3$N$_4$ structures is found to be $n=1.7552$ at a vacuum wavelength of $\lambda =$ \SI{660}{\nano\meter}. In the presented model only the fundamental modes of the waveguides are excited.
As it is indicated in Fig. \ref{fig:exportinterferencepatterns}(a), in order to estimate the periodicity of the intensity pattern, a sinusoidal function of the form $I(x) = \exp\left(-\left(\frac{x}{a}\right)^2\right) \sin^2\left(bx\right)$ has been fitted; the pattern spacing is found to be \SI{445}{\nano\meter}. The light propagating in the separate waveguides is among others defined by its phase, i.e. the position of the wave front. Manipulating the phases of the waveguides will introduce a shift in the interference pattern position, which is the phase shift needed for SIM. 

Figure \ref{fig:exportinterferencepatterns}(b) shows respective simulation results for counter-propagating waveguides, i.e. interfering at an angle of 180$^\circ$. As for the case of an interference angle of 50$^\circ$, the intensity distribution was fitted as $I(x) = \sin^2\left(ax\right)$ and a period of \SI{188}{\nano\meter} is found.

Although the pattern spacing can be calculated from using the effective mode index without running a full numerical simulation of the structure, here the simulation results are used to illustrate the expected intensity distribution over the whole waveguide geometry.

	\begin{figure}
		\centering
		\includegraphics[width=0.8\linewidth]{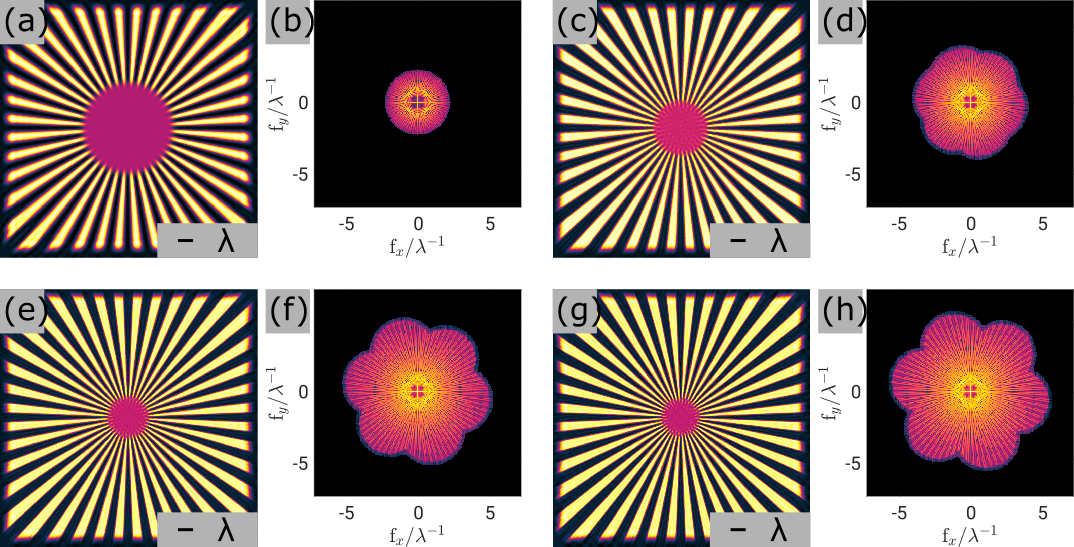}
		\caption[Simulate SIM N.A. = 1.2]{The resolution improvement using chip based SIM is simulated for an imaging objective with $\mathrm{N.A.} = 1.2$.
		(a) and (b) show the result of conventional deconvolution imaging at the indicated wavelength $\lambda$ and the representation in Fourer space respectively.
		(c) and (d) show the SIM result using interference patterns generated by two waveguides at an angle of \SI{60}{\degree}.
		(e-h) show the SIM resluts for patterns generated using an angle of \SI{120}{\degree} and \SI{180}{\degree} respectively.
		}
		\label{fig:SimulateChipSIM_NA_12}
	\end{figure}

 	\begin{figure}
		\centering
		\includegraphics[width=0.8\linewidth]{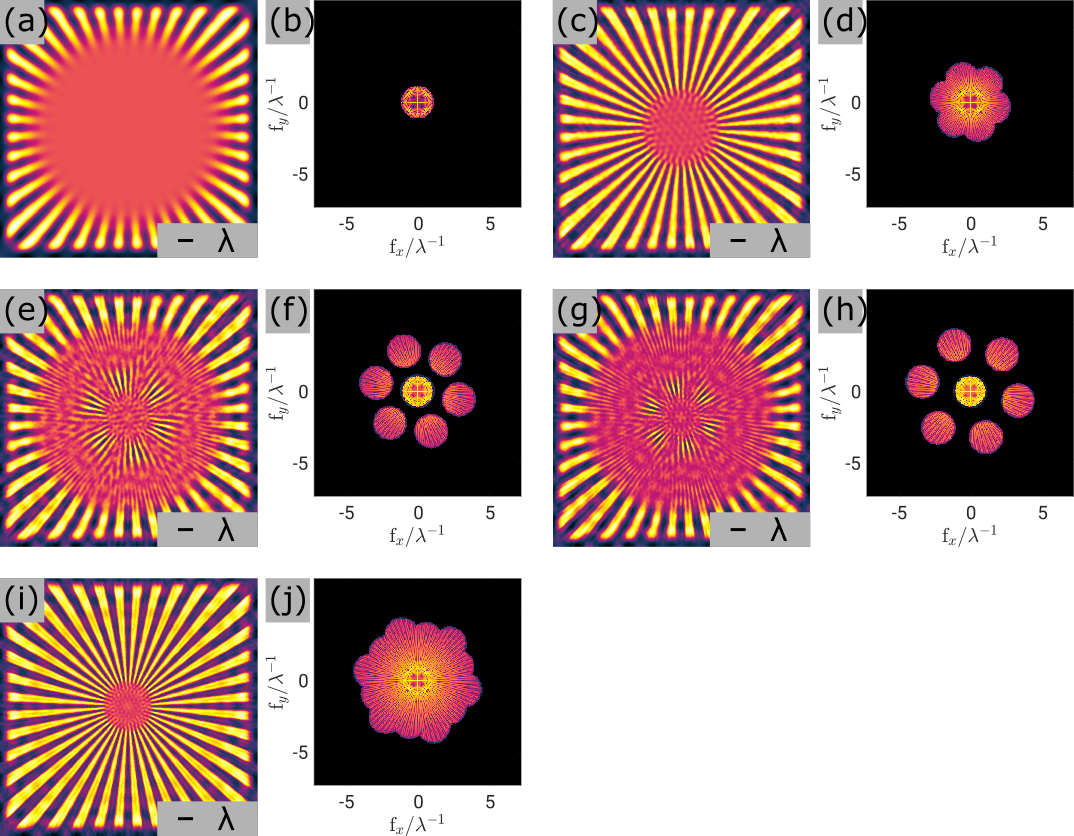}
		\caption[Fringes on waveguide]{
		The resolution improvement using chip based SIM is simulated for an imaging objective with $\mathrm{N.A.} = 0.6$.
		(a) and (b) show the result of conventional deconvolution imaging at the indicated wavelength $\lambda$ and the representation in Fourier space respectively.
	    (c) and (d) show the SIM results for patterns generated at an waveguide interference angle of \SI{60}{\degree}.
	    (e)-(f) and (g)-(h) show the SIM result using interference patterns generated at angles of \SI{120}{\degree} and \SI{180}{\degree} respectively.
	    In (i) and (j) the SIM reconstructions using all interference angles (\SIlist{60; 120; 180}{\degree}) are shown.
	    }
		\label{fig:SimulateChipSIM_NA_06}
	\end{figure}

The effect of structured illumination using planar waveguides is presented in Figs. \ref{fig:SimulateChipSIM_NA_12} and \ref{fig:SimulateChipSIM_NA_06} for two different imaging objective lenses using the Siemens star sample.

In Fig. \ref{fig:SimulateChipSIM_NA_12}, imaging using an objective lens with  $\mathrm{N.A.} = 1.2$ is simulated. Figure \ref{fig:SimulateChipSIM_NA_12}(a) shows the spatial result using plain illumination, and Fig. \ref{fig:SimulateChipSIM_NA_12}(b) shows the logarithmic values of the Fourier spectrum. It is visible in the figure that the highest frequency is limited to $2 \mathrm{N.A.}/\lambda = 2.4 \lambda^{-1}$. For a sample that is illuminated with a sinusoidal stripe pattern using planar waveguides with an effective refractive index of $n_\mathrm{eff} = 1.7$ at an angle between the waveguides of \SI{60}{\degree} the expected SIM result is shown in Fig. \ref{fig:SimulateChipSIM_NA_12}(d)-\ref{fig:SimulateChipSIM_NA_12}(e).
Increasing the interference angle to \SIlist{120;180}{\degree} increases the modulation frequency of the illumination pattern and thus the expected resolution improvement as shown in Figs. \ref{fig:SimulateChipSIM_NA_12}(e)-\ref{fig:SimulateChipSIM_NA_12}(f) and Figs. \ref{fig:SimulateChipSIM_NA_12}(g)-\ref{fig:SimulateChipSIM_NA_12}(h) respectively.

In a similar fashion cSIM imaging is simulated in Fig. \ref{fig:SimulateChipSIM_NA_06} for an objective lens with $\mathrm{N.A.} = 0.6$. Figures \ref{fig:SimulateChipSIM_NA_06}(a)-\ref{fig:SimulateChipSIM_NA_06}(b) show the result from using plain illumination. It shows a clear reduction in resolution as compared to the result using an imaging objective with a $\mathrm{N.A.} = 1.2$ as in Figs. \ref{fig:SimulateChipSIM_NA_12}(a)-\ref{fig:SimulateChipSIM_NA_12}(b).

Using an illumination pattern generated by waveguides with an interfering angle of \SI{30}{\degree} will improve the resolution as presented in Figs. \ref{fig:SimulateChipSIM_NA_06}(c)-\ref{fig:SimulateChipSIM_NA_06}(d).
For cSIM using patterns generated with interference angles of \SIlist{120; 180}{\degree} it is possible to obtain sample information of high spatial frequency as illustrated in Figs. \ref{fig:SimulateChipSIM_NA_06}(e)-\ref{fig:SimulateChipSIM_NA_06}(f) and Fig. \ref{fig:SimulateChipSIM_NA_06}(g)-\ref{fig:SimulateChipSIM_NA_06}(h) respectively.
Since the Fourier space is not filled evenly in this approach, a combined use of illumination patterns generated at interference angles of \SIlist{60; 120; 180}{\degree} as shown in Figs. \ref{fig:SimulateChipSIM_NA_06}(i)-\ref{fig:SimulateChipSIM_NA_06}(j) will improve the resolution without the loss of frequency bands. Thus, cSIM supports the use of low magnification/N.A. objective lens to enable super-resolution imaging over large field of view (FOV), while the maximum attainable resolution will be supported by the illumination provided by the waveguide as demonstrated in Figs.  \ref{fig:SimulateChipSIM_NA_06}(i)-\ref{fig:SimulateChipSIM_NA_06}(j). In conventional SIM, the illumination and the collection light paths are coupled, thus a use of low. N.A. objective lens to image large FOV will significantly reduce the supported optical resolution.

\section*{Materials and Methods}
\subsection*{Chip Design}
The key parameters in SIM are interference fringe spacing, phase modulation and fringe contrast. For chip-based SIM the dimensions of the active region is also crucial, to ensure that biological specimens are evenly illuminated by the structured illumination. The lateral size of cells are often from 10 $\mu$m and upwards. The waveguide platform has been developed using the high-refractive index material Si$_3$N$_4$. High-index contrast materials enable tight confinement of the light inside the waveguide which entails compact bend radius. This allows for ultra-compact and dense waveguide structures with small footprint benefiting from high intensity in the evanescent field. In the visible range, strong evanescent fields can be achieved by designing waveguides with core thickness between 100-200 nm. For a 150 nm thick core material; single mode condition, adiabatic taper condition and bending losses were previously investigated \cite{RN520}.
 
For this work single mode condition is necessary to generate uniform fringe pattern, which can be formed by interfering fundamental mode light guided inside the waveguide. The presence of any residual higher order modes will generate a mode-beat pattern which is undesirable. Thus, shallow rib waveguides were preferred over strip waveguides. For a shallow rib, fundamental mode guiding can be achieved using waveguide widths (1-1.5 $\mu$m wide) within the reach of fabrication using standard photo-lithography techniques. Waveguides with 4 nm rib height (total height of 150 nm) and around 1-1.5 $\mu$m width was adiabatic tapered out to 25 $\mu$m or 50 $\mu$m with a tapering length of 2 mm or 4 mm, respectively. A significantly low bending loss for a shallow rib waveguides was achieved for a bend radius of 2 mm and more. The optimization of the designed parameters can be found in previous literature \cite{RN520}.

\subsection*{Chip fabrication}

\begin{figure}[h]
    \centering
    \includegraphics[width=.4\linewidth]{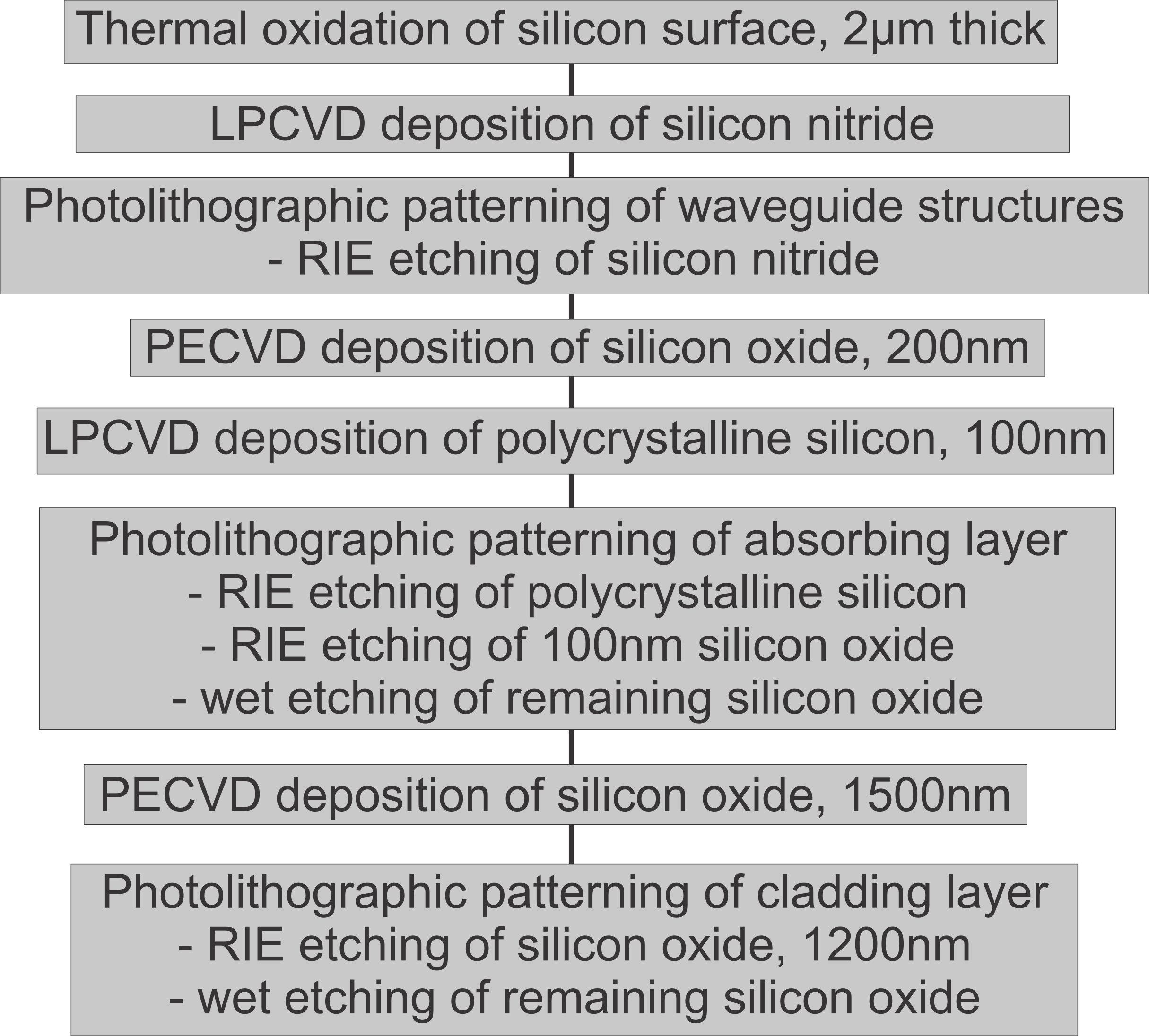}
    \caption{Flow-chart showing the fabrication steps performed in the production of cSIM waveguide chips}
    \label{workflow}
\end{figure}

The production of waveguide chips was performed at the Insitute of Microelectronics Barcelona (IMB-CNM, Spain). First, an oxide layer of ca. 2 $\mu$m was thermally grown on a silicon slab, followed by low-pressure chemical vapour deposition (LPCVD) of 150 nm Si$_3$N$_4$ at 800$^\circ$. Conventional photo-lithography imprinted the waveguide geometries on a layer of photo-resist. 4nm rib waveguides were realized using reactive-ion-etching (RIE). In order to prevent cross-talk of light into adjacent waveguide structures, an absorption layer consisting of 200 nm SiO$_2$ (deposited by plasma-enhanced chemical vapour deposition (PECVD)) followed by 100 nm poly-crystalline silicon (deposited by PECVD at 300$^\circ$) was designed as a negative image of the first mask with 10 $\mu$m added to the waveguide width. RIE was used to etch until 100 nm of SiO$_2$ remained, with wet etching (using hydrofluoric acid) removing the remaining material from the waveguide surface. Finally, a 1.5 $\mu$m SiO$_2$ layer (deposited by PECVD) built the cladding layer. The chip fabrication workflow is shown in Fig. \ref{workflow}.  

 Imaging areas were patterned by a 3rd photo-lithography step. The window openings were realized using RIE until 100 nm SiO$_2$ was remaining, followed by wet etching to completely remove the oxide layer while preserving the Si$_3$N$_4$. Further details of the fabrication optimization and process can be found elsewhere \cite{RN682}.
 
 \subsubsection*{Deposition of thermo-optical heating element}
 \label{thermosection}
\begin{figure}[h]
     \centering
     \includegraphics[width=0.8\linewidth]{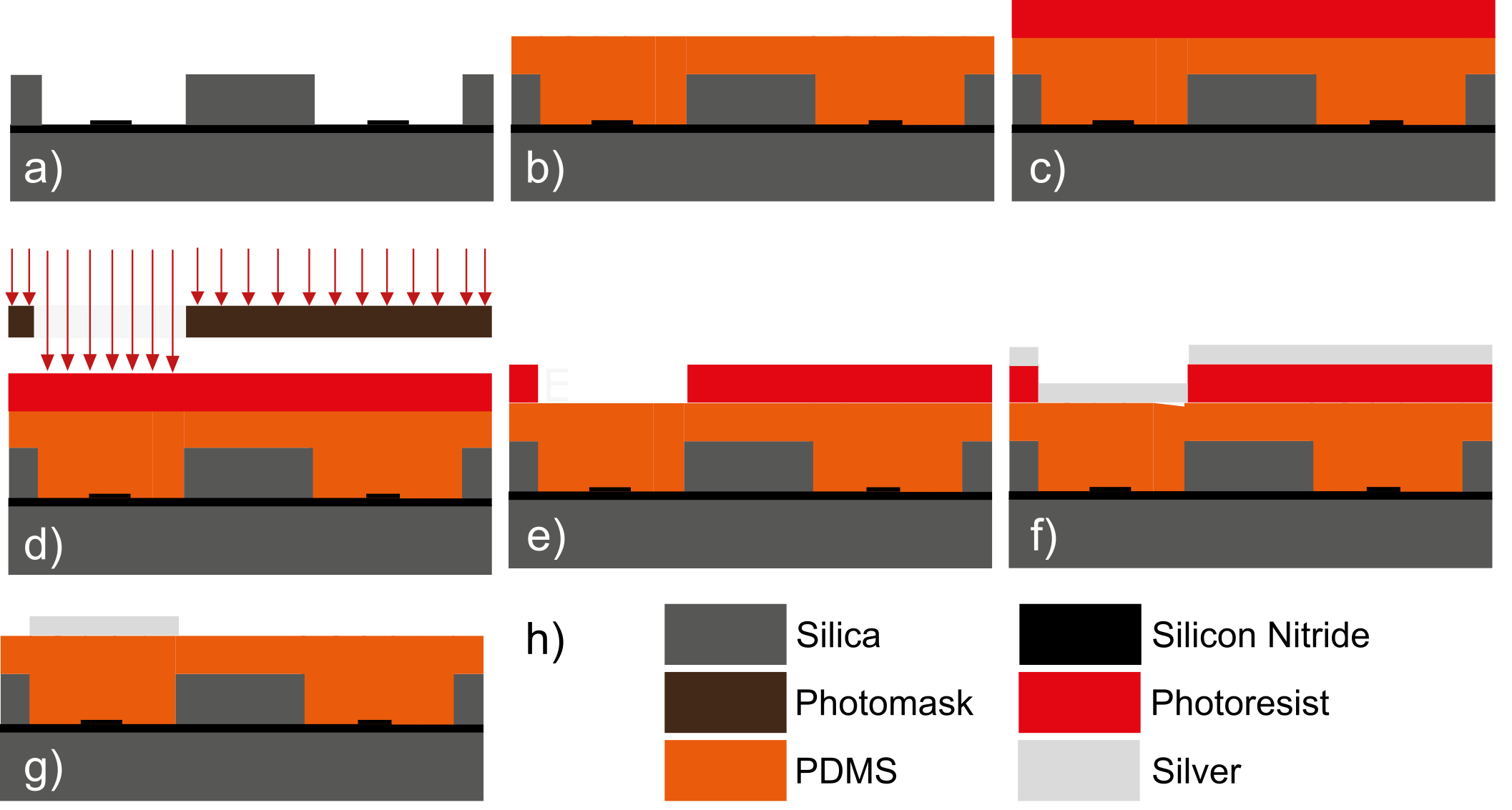}
     \caption{Fabrication of thermo-electro-opitcal heating element. (a) Both arms of the cSIM structure incorporates opening in the cladding. (b) A cladding layer made from  either PDMS or Su-8 covers the entire chip, followed by a photo-resist (c). (d) One of the openings is exposed and the photo-resist was removed. (e) A layer of Ag is sputtered on the entire chip, before the photo-resist is removed leaving only silver on one of the arms.}
     \label{LiftOff}
 \end{figure}

 The cSIM phase-stepping using the design depicted in Fig.\ref{Designs}(c) was achieved using a resistive heating element positioned on one arm for each of the SIM angles. Silver resistors were sputtered on a cladding layer of either PDMS or Su-8, followed by a photo-lithography step to remove residue silver not attached at the window openings as depicted in Fig. \ref{LiftOff}. Electrical contact to the sputtered circuit was made using silver epoxy.

\subsection*{Experimental setup}
\begin{figure}
    \centering
    \includegraphics[width=1.0\linewidth]{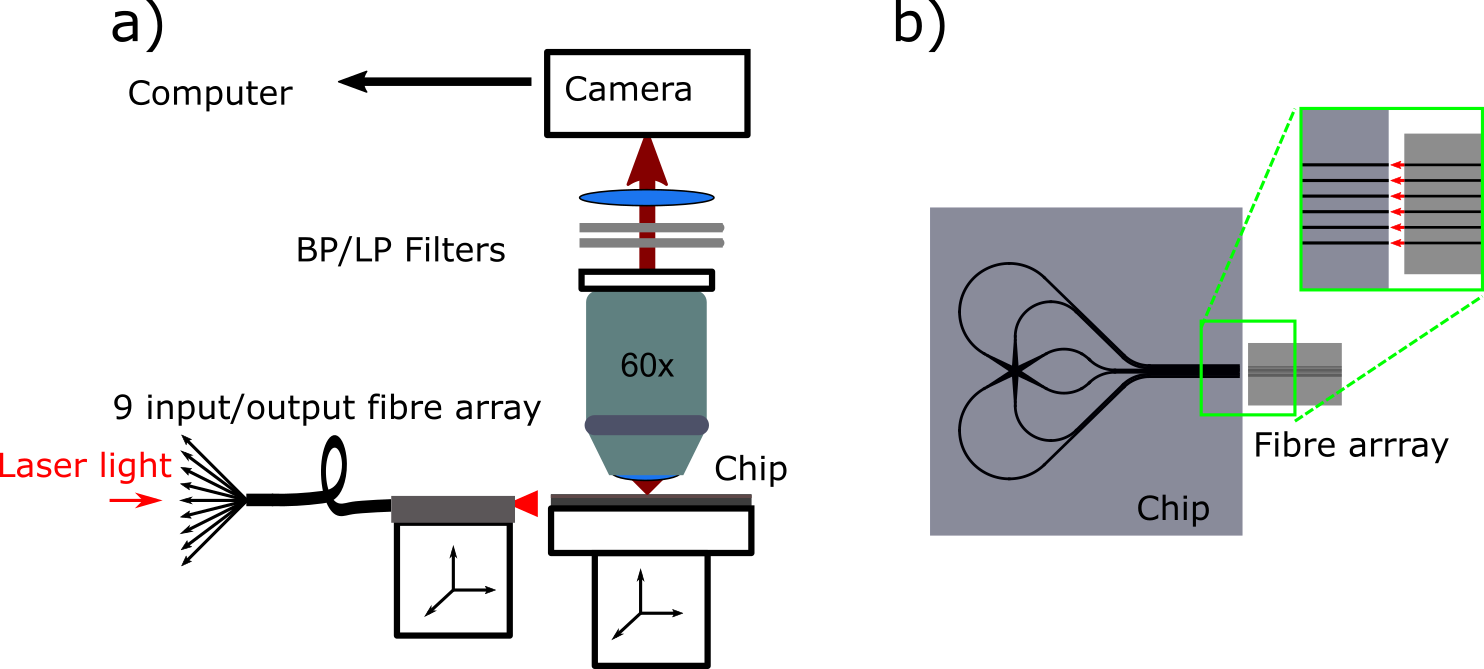}
    \caption{Experimental setup. a) An upright microscope detects the emitted fluorescence. The waveguide chip is held by a vacuum chuck while light is coupled on to the chip via a nine-fibre array adapter. Both sample and coupling stages have xyz movement. b) The fibre array adapter holds 9 optical fibers with equidistant spacing (127 $\mu$m), which corresponds to the waveguide spacing on the chip. Each of the 9 optical fibers can be individually activated, according to the chip designs being used (Fig. \ref{Designs}).}
    \label{setup}
\end{figure}

A microscope modified from an Olympus modular microscope (BXFM) was mounted on a XY motorized translation stage. The microscope was fitted with a Hamamatsu Orca flash sCMOS camera. cSIM images were acquired using either Olympus x60/1.3SiO or Olympus x60/1.2W objective lenses.  The waveguide chip was mounted on a micrometer xyz-stage using a vacuum chuck to hold the chip. Light was coupled on to the chip using a nine-fiber array adapter, as seen in figure \ref{setup}(a). The fibre array adapters are commercially available with fixed spacing (e.g. 127 $\mu$m). The optical waveguides were separated by the same space (127 $\mu$mm) as the fibre array, allowing for multiple waveguides to have light coupled at the same time, as seen in Fig.\ref{setup}(b). For the design depicted in Fig. \ref{Designs}(b), an additional 50:50 fused fibre-split (OZ optics) was used to split the light. To allow phase stepping using the design depicted in Fig. \ref{Designs}(c), a voltage supply was coupled to the resistive on-chip electro-optical heating elements. The image data was analyzed using Fiji open source image processing software and FairSIM \cite{RN666}, an open source SIM reconstruction plugin. cSIM data from the \SI{180}{\degree} interference waveguides was analyzed with custom software.

\subsection*{\textit{d}STORM field mapping}

The fringe period formed by waveguide design with an interfering angle of 180$^\circ$ was beyond the resolution limit of the imaging objective lens. Therefore, {d}STORM near field mapping was performed by staining the waveguide surface with a layer of fluorescent molecules. A 1/1000 solution of CellMask deep red (Thermofisher, C10046) in phosphate buffered saline (PBS) was incubated for 20 minutes on the waveguide surface at room temperature. The solution was aspirated and a PDMS micro-chamber positioned on top of the chip was filled with \textit{d}STORM switching buffer consisting of an enzymatic oxygen scavenging system and 100 mM $\beta$-Mercaptoethylamine (MEA). The surface was excited with the 660 nm laser with sufficiently high power to achieve photo-switching. A stack of 100000 images was acquired at 30 ms exposure time and reconstructed using the Fiji plugin ThunderStorm \cite{RN484}.

\section*{Experiments and Results}

\begin{figure}
    \centering
    \includegraphics{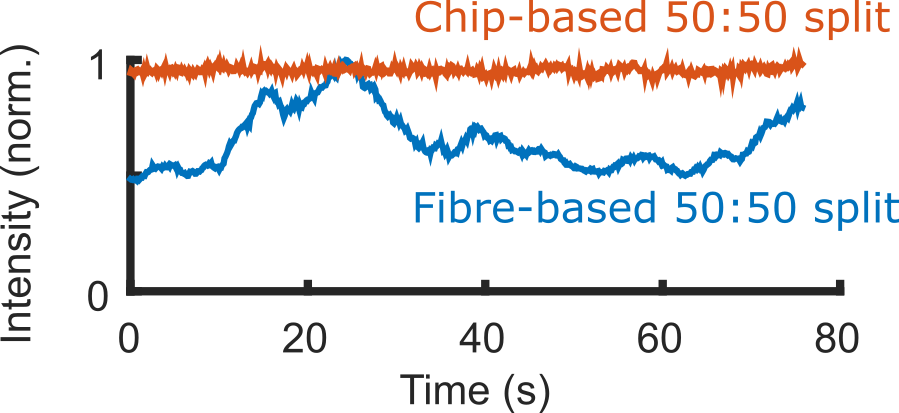}
    \caption{The phase stability of the standing wave SIM interference pattern is shown indirectly by tracking the fluorescence from a bead over time. The blue curve shows the measured bead intensity using the waveguide structure depicted in Fig. \ref{Designs}(b) splitting the light off-chip using a 50:50 fibre split, while the red curve show the detected intensity for the approach depicted on Fig. \ref{Designs}(c) splitting the light on-chip using a y-branch. The unstable blue curve show that the phase is drifting in time when using the fibre-split, while splitting the light on the chip yields a stable phase.}
    \label{Phasestability}
\end{figure}

The modulation index for the standing wave interference pattern, created by the cSIM waveguide, was measured using the signal from fluorescent beads with a diameter of 100 nm. The waveguide interference angle was 50$^\circ$ which corresponds to around 460 nm fringe spacing (eq. \ref{fringespaceing}). The intensity of the beads was tracked over time while the phase of the interference pattern was changing slowly. The bead intensity was observed going from bright at high fringe amplitude to dim at low fringe amplitude. This amplitude variation was used to calculate the modulation index $\nu$  as $\nu=\frac{I_{max}-I_{min}}{I_{max}+I_{min}}$ and found to be around 78\%. Furthermore, the phase stability of the cSIM pattern over time was measured by looking at a fluorescent bead sample without changing the phase of the cSIM pattern. The measurement was carried out for the structures depicted in Figs. \ref{Designs}(b)-\ref{Designs}(c), to compare the effect on phase stability between splitting the light using a y-branch on the chip, or using the fibre-split on the optical table. In Fig. \ref{Phasestability} the normalized fluorescence signal from a bead tracked over 80s is shown. The chip-based y-branch split results in a stable phase measurement, while for the fiber version the phase clearly drifts over time.

\begin{figure}
    \centering
    \includegraphics[height=3.5cm]{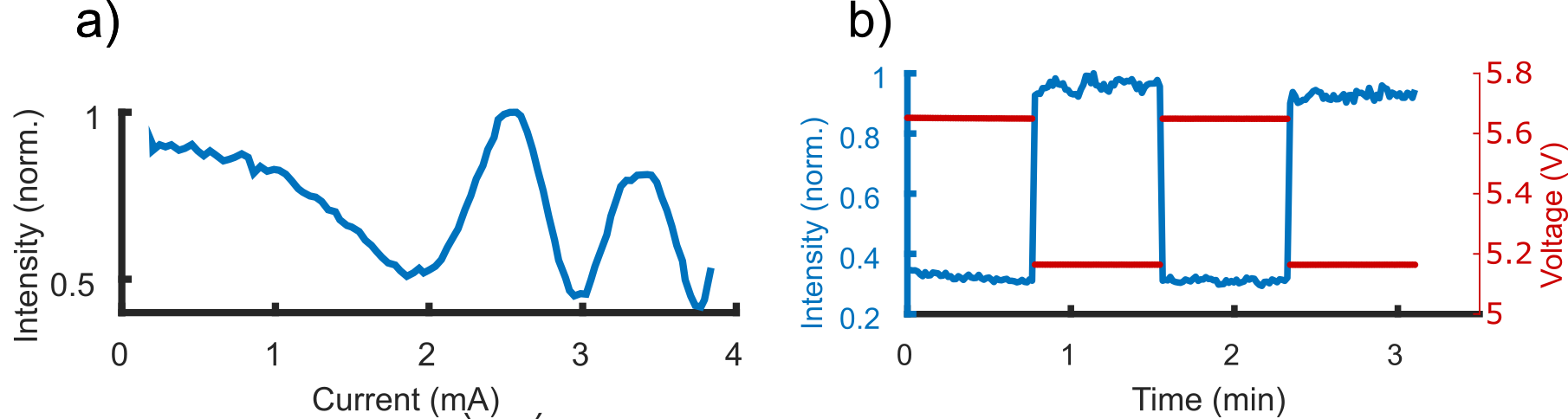}
    \caption{a) By applying current to the thermo-optical element the phase of the SIM interference pattern was changed. A fluorescent bead was tracked while changing the current, thus shifting the SIM pattern phase. b) By selecting the voltages  corresponding to a $\pi$ change in the SIM pattern, the phase was stepped between 0 and $\pi$ over 3 minutes to demonstrate the reproducibility of the phase stepping method. }
    \label{Pichange}
\end{figure}

For successful cSIM imaging, equidistant phase stepping will need to be reproduced over time. One approach to achive this is by using electro-optical heating elements positioned on one of the arms for the designs depicted in Fig.\ref{Designs}(c).  Fig. \ref{Pichange}(a) shows how the phase of the interference pattern increases with current flowing in the electro-optical heating element. This is once again indirectly observed as the oscillation of the intensity measurement from the fluorescent beads.  From the curve in Fig. \ref{Pichange}(a), the voltage corresponding to a phase shift of $\pi$ was found stepping between 5.17 V and 5.65 V. To showcase that phase stepping can be reproduced, we let the voltage jump back and fourth over time giving a $\pi$ shift as seen if Fig. \ref{Pichange}(b).

\begin{figure}
\centering
  \includegraphics[width=1.0\linewidth]{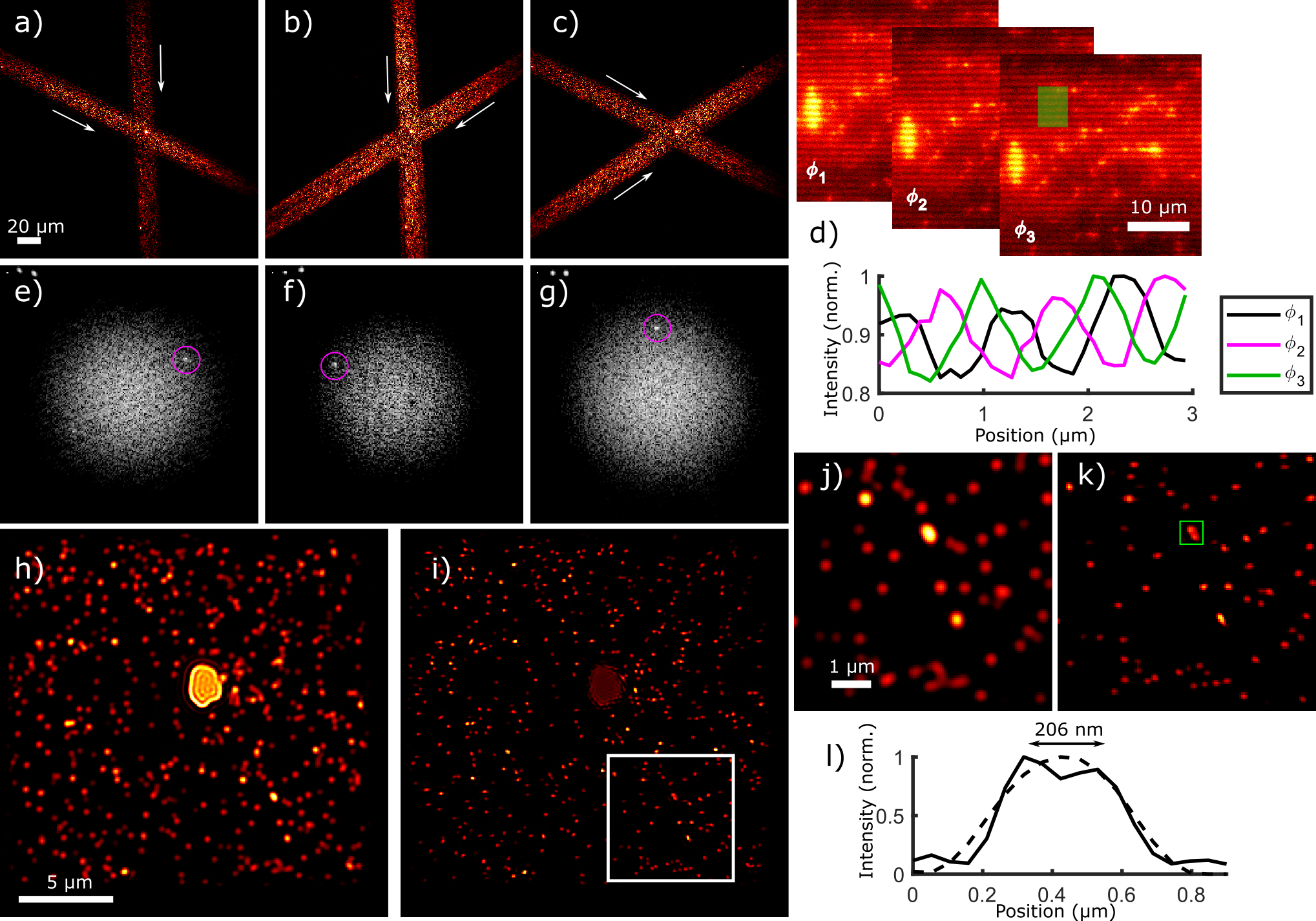}
  \caption{Demonstration of cSIM imaging using a sample of fluorescent beads and the waveguide structure shown in Fig. \ref{Designs}(b) interfering at 60$^\circ$ angle,. (a-c) Light is interfering in the overlapping regions coming from 3 different rotational angles. (d) Each of the rotations generate an interference pattern. The pattern is phase-stepped three times, as seen in the upper panel. The line-profile is indicated with a green box, and show the equidistant phase-steps needed for each cSIM rotational angle. Note that (d) is taken from a waveguide interfering at 20$^\circ$. This is to demonstrate the phase-stepping with high contrast fringes. (e-g) show the peaks in Fourier space showing interference fringe spacing and rotations. Here, the peaks in (e) corresponds the active waveguides shown in (a), (f) corresponds to (b) and (g) corresponds to (c). In the overlapping region the 9 images are reconstructed into one SIM image. (h) show the diffraction limited and (i) show the cSIM reconstruction. (j-k) show the zoom indicated in (i) and (l) the line-profile indicated in (k). Two beads spaced 206 nm apart is clearly separated in the cSIM image, but is not resolved in the diffraction limited image. Excitation wavelength of 660 nm and Olympus x60/1.3SiO imaging objective lens used in this experiment.} 
  \label{OffchipResult}
\end{figure}

To demonstrate cSIM super-resolution imaging, a sample of 100 nm large fluorescent beads was imaged using the design depicted in Fig \ref{Designs}(b), lower panel. The sample was placed on the chip within the imaging area and laser light was sequentially coupled on to the three SIM angles. Figures \ref{OffchipResult}(a)-\ref{OffchipResult}(c) show frames from the raw data where the active waveguides are visible via the fluorescent bead layer. The direction of light propagation is indicated with white arrows. Fig. \ref{OffchipResult}(d) shows three images with same pattern orientation, but with different phase steps. Equidistant phase-steps was extracted from the data as is shown in the plot of the line-profile marked with green color. These images are from an experiment using 20$^\circ$ interference angle, to clearly visualize the presence of the standing wave using the high contrast fringes achieved with a low interference angle. The orientation of the patterns shown in Fig. \ref{OffchipResult}(d) corresponds to the interference shown in Fig. \ref{OffchipResult}(c). Figures \ref{OffchipResult}(e)-\ref{OffchipResult}(g) show the cross-correlation used as an intermediate result by the FairSIM reconstruction software with magenta circles marking the detected modulation frequency and pattern orientation. The orientation and detected frequency in Figs. \ref{OffchipResult}(e)-\ref{OffchipResult}(g) corresponds with the angle of interference in Figs. \ref{OffchipResult}(a)-\ref{OffchipResult}(c) respectively. The overlapping area of Figs.\ref{OffchipResult}(a)-\ref{OffchipResult}(c) was cropped and reconstructed using FairSIM. Figures \ref{OffchipResult}(h)-\ref{OffchipResult}(i) show the filtered diffraction limited image and the cSIM reconstruction respectively, with a clear resolution enhancement visible. A cluster of beads saturates the image near the center, however it does not impact the reconstruction. The zoomed images in Figs. \ref{OffchipResult}(j)-\ref{OffchipResult}(k), marked with a white box in Fig. \ref{OffchipResult}(i), further visualize the resolution enhancement, and  Fig. \ref{OffchipResult}(l) shows a line-profile, marked with a green box in Fig. \ref{OffchipResult}(k), measuring the distance between two beads. The beads are separated by 206 nm, which is clearly resolved in the cSIM image while not resolved in the diffraction limited image. 

\begin{figure}[h]
    \centering
    \includegraphics[width=0.8\linewidth]{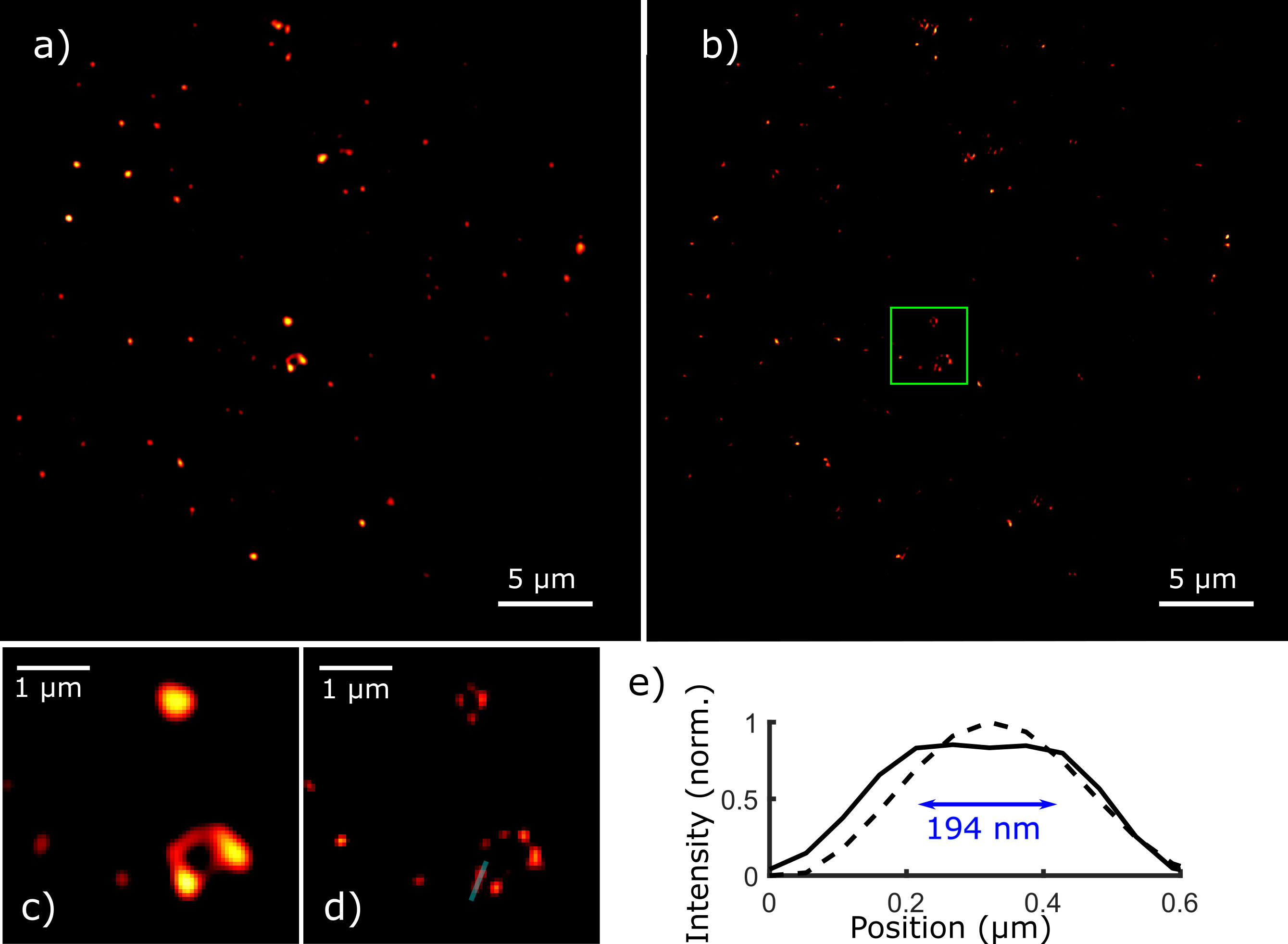}
    \caption{Chip-based SIM proof of principle using interference at 50$^\circ$ and thermo-optical phase shifting with the structure depicted in Fig.\ref{Designs}(c). The sample is 100 nm fluorescent beads imaged with evanescent excitation light  using the waveguide chip. (a) Show the diffraction limited image while (b) show cSIM reconstruction. (c) and (d) show the zoom indicated with a green box in (a) and (b).  (e) a line-profile show that 2 beads laying 217 nm appart are un-resolved in the diffraction limited image (dotted line), while the cSIM reconstructed image the beads are resolved. Excitation wavelength of 660 nm and Olympus x60/1.2W imaging objective lens used in this experiment.}
    \label{ThermoSimresult}
\end{figure}

Furthermore, on-chip phase stepping using thermo-optical phase modulation using the structure shown in Fig. \ref{Designs}(c), lower panel, was used to generate a cSIM image of fluorescent beads. The chip was prepared using Su-8 with silver circuits as the thermo-optical layer. The phase change was achieved by changing the voltage thus creating heat in the window openings allowing for thermal expansion of the Su-8 and consequently imparting a change in the phase. Figures \ref{ThermoSimresult} (a)-\ref{ThermoSimresult}(b) show the diffraction limited image and cSIM reconstruction respectively. Figures \ref{ThermoSimresult}(c)-\ref{ThermoSimresult}(e) show the zoomed image, marked in green and a line-profile showing two beads separated by 194 nm resolved in the cSIM image.

\begin{figure}[h]
    \centering
    \includegraphics[width=0.8\linewidth]{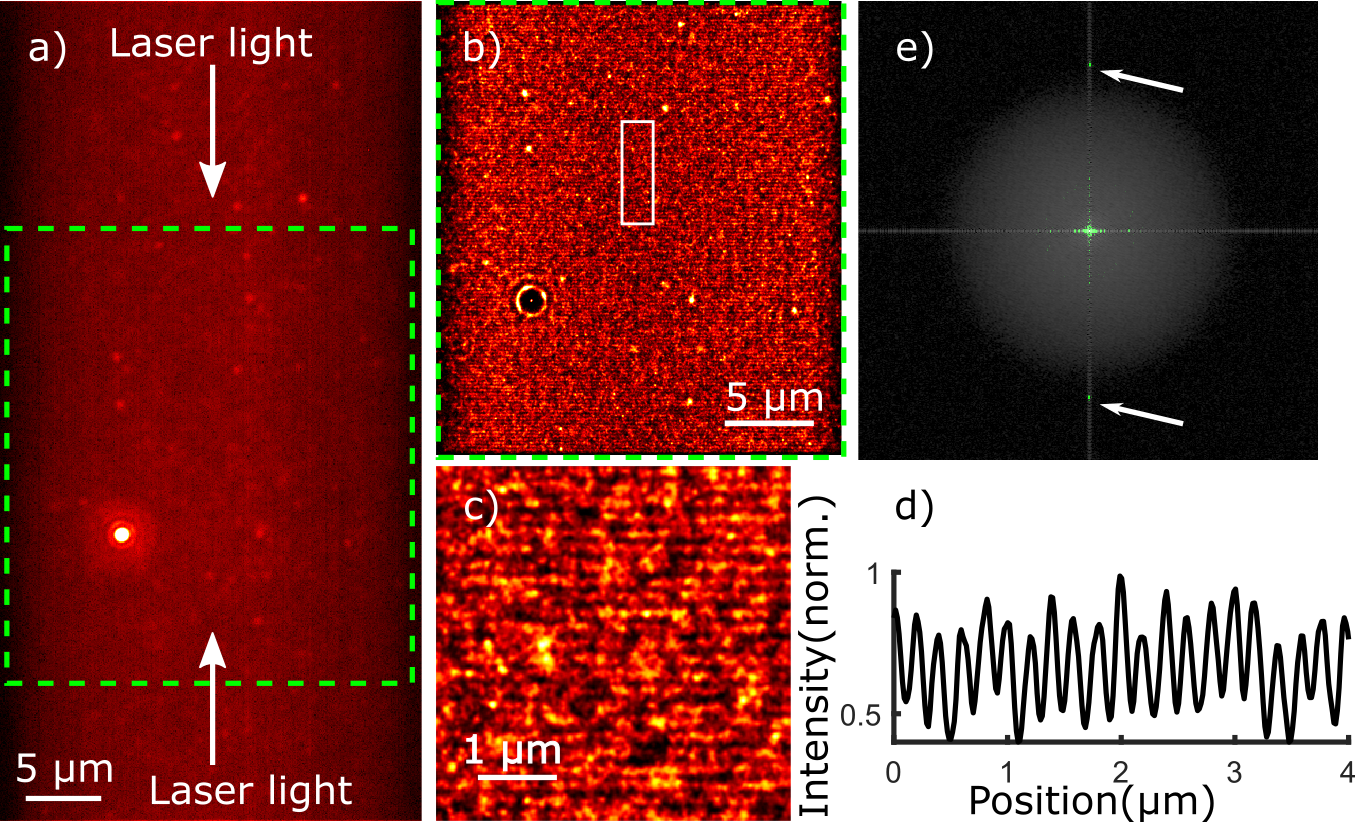}
    \caption{Visualizing a diffraction limited standing wave. (a) Counter propagating light excites fluorescence at the interference region using the structure depicted in Fig.\ref{Designs}(a), top panel. The detected fluorescence emission does not indicate the presence of a standing wave due to the sub-diffraction fringe spacing. (b) Shows the \textit{d}STORM reconstruction of the region marked with a green dotted box in (a). The standing wave becomes visible as seen in the zoomed image in (c). (d) Shows the standing wave with a line-plot, marked with a white box in (b). (e) The power-spectrum of (a) is shown in gray color with the power spectrum of the \textit{d}STORM reconstruction in (b) overlaid in green color. The peaks marked with arrows show the presence of the standing wave, and its location outside the cutoff using diffraction limited microscopy.} 
    \label{dstormpattern}
\end{figure}

\begin{figure}[h]
    \centering
    \includegraphics[width=1.0\linewidth]{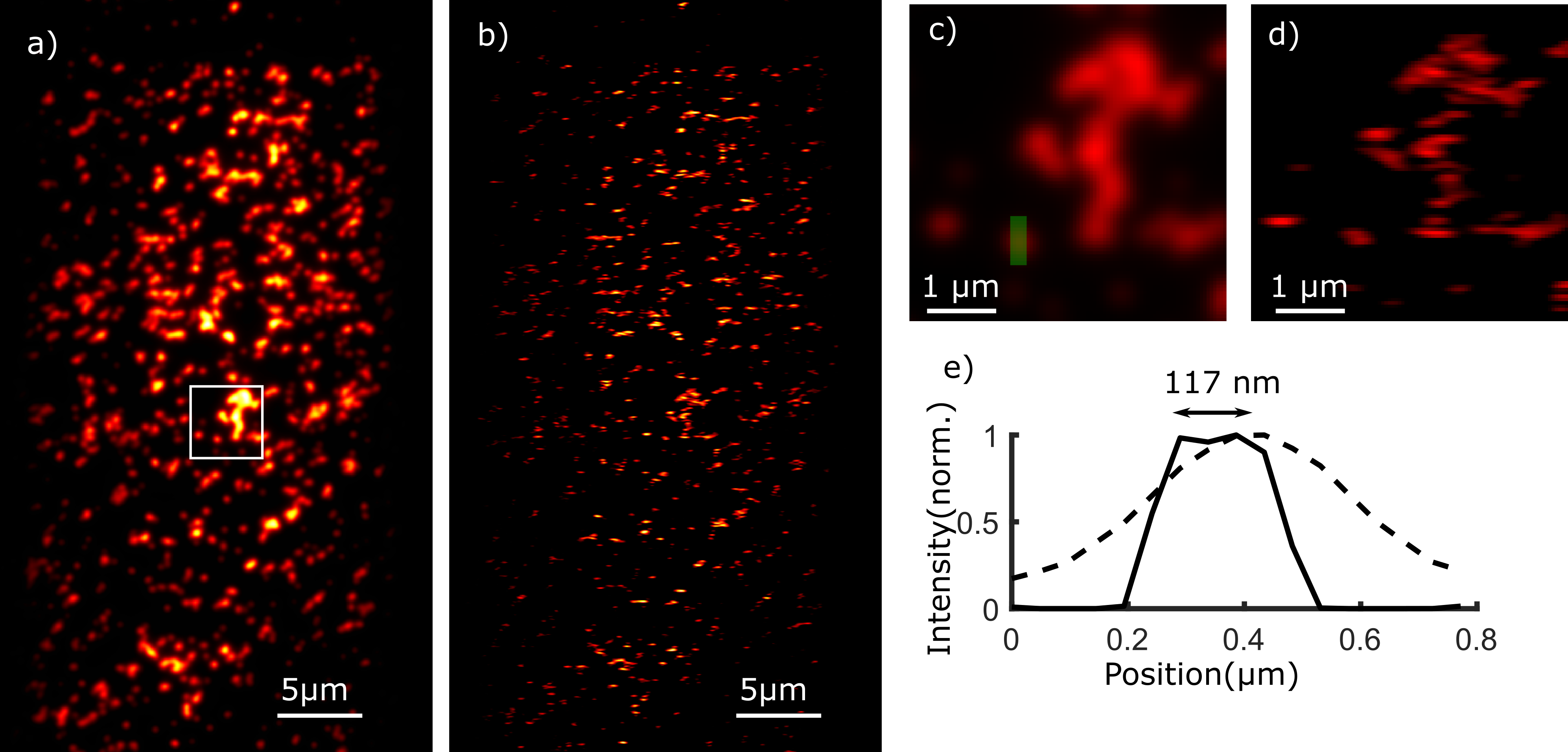}
    \caption{cSIM imaging using 180$^\circ$ interference angle. (a) A 100 nm bead sample is imaged using diffraction limited chip-based imaging using the structure depicted in Fig. \ref{Designs}(a), top panel. (b) Show the cSIM reconstruction with (c) and (d) showing the zoom indicated with a white box in (a). (e) A line-profile show two beads separated by 117 nm are resolved using cSIM (solid line), and not resolved in the diffraction limited image (dotted line). The resolution enhancement is around 2.3x surpassing what is available in conventional SIM. Excitation wavelength of 660 nm and Olympus x60/1.2W was used in this experiment}
    \label{1DSim}
\end{figure}

The ultimate goal of cSIM is to surpass the resolution enhancement of conventional linear SIM, which can be achieved using opposing pair of waveguides at 180$^\circ$. This will generate interference fringe with smaller spacing than what is presently possible using a high N.A. objective lens. Such chip design would enable a resolution enhancement beyond a factor of 2x (as in case of linear SIM) and using silicon nitride waveguide platform and an imaging objective lens of 1.2 N.A. the theoretical resolution enhancement of 2.4x (eq. \ref{resolution}) is possible. 1D counter-propagating waveguides as depicted in Fig. \ref{Designs}(a), top panel, was used to investigate this idea. To verify the presence of a standing wave, the surface of the chip was stained using a fluorescent dye. The emitted fluorescence was captured by the camera, with no indication of a standing wave visible as seen in Fig. \ref{dstormpattern}(a) using a 1.2 N.A objective lens. By dSTORM imaging, the fluorescent dye molecules were photo-switching allowing them to be localized at a precision high enough to visualize the underlying standing wave as seen from the dSTORM reconstruction in Fig. \ref{dstormpattern}(b). Figure \ref{dstormpattern}(c) show the zoom marked with a box in Fig. \ref{dstormpattern}(b) where the standing wave pattern is clearly visible and further backed up with the line-profile in Fig. \ref{dstormpattern}(d), taken from the region marked in white in Fig. \ref{dstormpattern}(b). Figure \ref{dstormpattern}(e) shows the power-spectrum of Fig.\ref{dstormpattern}(a) in gray color overlaid with the power-spectrum of Fig. \ref{dstormpattern}(b) in green color. The peaks marked with arrow show the frequency and orientation of the pattern, and its position outside the cutoff of the diffraction limited image (i.e Fig. \ref{dstormpattern}(a)). The same structure was used to generate a 1D cSIM image as shown in Fig. \ref{1DSim}. The diffraction limited image is shown in Fig. \ref{1DSim} (a) and the 1D cSIM reconstruction i Fig. \ref{1DSim}(b). Figures \ref{1DSim}(c)-\ref{1DSim}(d) show the zoom marked with a white box in Fig. \ref{1DSim}(a). The line-profile shows two beads separated by 117 nm is separated in the cSIM image, which yields a 2.3x resolution enhancement over the Abbe resolution limit supported by the imaging objective lens of 1.2 N.A at 660 nm.

\section*{Discussion}

Conventional structured illumination microscopy is gaining its popularity as the preferred optical nanoscopy method, owning to its ease of use, compatibility with most of the bright and stable fluorophores, high imaging speed and its usability in live cell imaging applications. However, the limitation of the present SIM solution is high-cost and high-complexity which has hindered its widespread penetration. Present day SIM employ bulk optics for laser beam engineering in free-space and uses a microscope objective lens to deliver the illumination pattern to the sample. This approach is prone to misalignment and therefore, successful implementation requires well-calibrated optics hosted in a stable and mechanically rigid platform, consequently increasing the cost and the complexity of the method. In this work we have demonstrated a new concept of performing TIRF-SIM using a photonic chip. The proposed cSIM can simplify the optical configuration enabling super-resolution SIM imaging using any standard optical microscope, by retrofitting the proposed photonic chip. The entire illumination light path, i.e. light delivery, pattern generation and beam steering is provided using the photonic chip, therefore any standard optical microscope can be used to generate super-resolution SIM images. The proposed experimental setup, together with on-chip phase splitting is inherently very stable with minimum bulk optical elements used and therefore free-from miss-alignment issues. Moreover, the waveguide chip can be fibre-pigtailed with an optical fibre for additional stability and integration. The compatibility of a photonic-chip with optical fibres enable easy wavelength multiplexing, where adding or removing optical components does not miss-align the system. 

Furthermore, cSIM has the potential to further extend the resolution provided by conventional SIM. In conventional SIM the illumination pattern is generated through the imaging objective, where the pattern frequency depends on the N.A. of the objective lens and the excitation wavelength. The resolution enhancement of conventional linear SIM is thus limited to a factor of 2x. On the other hand, in cSIM the illumination fringes are generated by the photonic chip and does not rely on the N.A. of the imaging/collection objective lens. The illumination and the collection light paths are completely decoupled. It was demonstrated that by using a waveguide material of high-refractive index, the fringe spacing of chip-based SIM will be smaller than what can be generated using the high N.A oil-immersion objective lens commonly used in conventional SIM. In Fig. \ref{fig:SimulateChipSIM_NA_12}, the resolution improvement by using 180 $^\circ$ interference angles theoretically corresponds to around 20 \% enhancement over conventional SIM (depending on waveguide material as seen in eq.\ref{resolution}). We have experimentally verified this by using the 1D SIM structures from Fig.\ref{Designs}(a), top panel, as seen in Figs. \ref{dstormpattern}-\ref{1DSim}. Moreover, since the cSIM approach has decoupled the generation of the interference fringe pattern from the objective lens, a low N.A. and low magnification objective lens providing a larger FOV can be used while still maintaining the resolution improvement of the waveguide generated illumination pattern as is demonstrated in Fig. \ref{fig:SimulateChipSIM_NA_06}(i)-\ref{fig:SimulateChipSIM_NA_06}(j). This is an improvement over conventional SIM, where a use of low. N.A. objective lens to image large FOV will significantly reduce the supported optical resolution. 
The incomplete coverage of the frequency space as seen in Fig. \ref{fig:SimulateChipSIM_NA_06}(e)-\ref{fig:SimulateChipSIM_NA_06}(f) and Fig. \ref{fig:SimulateChipSIM_NA_06}(g)-\ref{fig:SimulateChipSIM_NA_06}(h) actually represent a filter in the frequency domain, only acquiring a well-defined set of higher spatial frequencies. This interesting property of cSIM (selective spatial filtering) might find potential usage in the future.

In this study, we investigated different waveguide geometries for performing SIM imaging. The differences in the geometries is mainly on how the phase stepping for the SIM pattern is achieved and where the light is split. For phase modulation, we propose three different strategies each having its own advantages and disadvantages.  First, off-chip phase modulation using two waveguide arms interfering at an angle and two waveguide arms excited simultaneously using two input beams as shown in Fig. \ref{Designs}(b). Here, the light is split using a 50:50 fibre-split and the phase modulation is achieved by changing the phase of one waveguide arm relative to the other, e.g by heating. This configuration is unstable due to environmental drifts, as observed in Fig. \ref{Phasestability}, but cSIM imaging could be performed by letting the camera run, recording a stack of images, while introducing a controlled phase-drift. Then the images corresponding to equidistant phase steps could be identified afterwards, before reconstructing the cSIM image. Obviously, this method lacks both speed and repeatability, making it unsuited for live-cell imaging. By splitting the light on-chip, the environmental phase instability was eliminated as seen in Fig. \ref{Phasestability}. For on-chip light splitting we investigated two strategies for phase-control. In Fig. \ref{Designs}(a), phase shifts are provided by activating different waveguide arms. Each input waveguide  has a Y-splitter creating a loop, and each of the waveguides have different pre-defined path-lengths, creating phase shifts relative to the other input waveguides. This method is the easiest to implement since no additional fabriaction is needed, and holds the potential for fast cSIM imaging since the phase-shifts only rely on coupling the light in to nine different waveguides (in case of 2D cSIM). This can be achieved, e.g. by using a fibre switch-box with at least nine outputs. In Fig. \ref{Designs}(c) on-chip dynamic phase modulation using thermo-optical phase modulation strategy is shown. This configuration also has the potential to do fast phase-shifting, since the phase relies on heating and cooling of a layer of Su-8 or PDMS. The rise and fall time for this process should be in the microsecond range, which would make the camera the limiting factor when it comes to image speed. However, heating via the silver circuits have shown to induce some thermal drift of the sample, which would not only shift the xy-position of the camera, but also induce loss of coupling. This problem should be alleviated by creating a custom sample/coupling stage with thermal cooling, and also by reducing the size of the sputtered silver circuited to a minimum so that the applied heat is more concentrated.

This is the first report of photonic-chip based SIM and some challenges should be beter addressed in future work. Firstly, the light guidance in the two interfering waveguide arms must be equal for high visibility. It was observed that any loss of power inside one of the waveguide arms result in decreasing the visibility of the fringes. The loss of power could be to the hot-spot, imperfections in the waveguides, or losses related to the PDMS/Su-8 deposition in the phase-change openings. For the best structures we could achieve fringe visibility of around 80 $\%$. On the other hand, if there are small losses in one of the arms, the visibility is greatly reduced. This can be seen in Fig. \ref{Pichange}(a). The phase is not affected by this, but the SIM reconstruction could fail if the visibility becomes to low. 

In future work, development of 2D SIM with 2.3-2.4X resolution enhancement will be studied by designing suitable waveguide geometries. The proposed cSIM should support high-speed imaging as it does not have any movable optical or mechanical parts, which will be investigated in the future work.  Although, the focus of this work is to report chip-based SIM, it can be foreseen that integrated photonic chips provides new research directions of performing complex light beam shaping for different advanced optical microscopy methodologies. It is also possible to develop even new illumination strategies as the waveguide chip allows easy manipulation of light using an integrated optics approach as compared to free-space optics. This could open avenues for using photonic-chips for  biological applications.

\nolinenumbers
\section*{Acknowledgement}
This work is funded by ERC starting grant (336716). This work has made use of the Spanish ICTS Network MICRONANO FABS partially supported by MEINCOM.
The authors would like to acknowledge Rainer Heintzman for help with reconstruction algorithms, Olav G. Hellesø for help with theory and simulations and Sourouv Das for help in the development of thermo-optical phase modulators.

\bibliography{library}

\bibliographystyle{unsrt}

\end{document}